\documentclass[%
 reprint,
superscriptaddress,
showpacs,
 amsmath,amssymb,
 aps,
prb
]{revtex4-1}

\usepackage{graphicx}
\usepackage{dcolumn}
\usepackage{bm}

\usepackage{color}

\usepackage{ulem} 

\begin{document}


\title{Orbital-selective electronic excitations in iron arsenides revealed by simulated nonresonant inelastic x-ray scattering}

\author{Kenji Tsutsui}
\affiliation{Quantum Beam Science Center, Japan Atomic Energy Agency, Hyogo 679-5148, Japan}

\author{Eiji Kaneshita}
\affiliation{Sendai National College of Technology, Sendai 989-3128, Japan}

\author{Takami Tohyama}
\affiliation{Department of Applied Physics, Tokyo University of Science, Tokyo 125-8585, Japan}


\date{\today}
             

\begin{abstract}
Nonresonant inelastic x-ray scattering (NIXS) is a possible tool to detect charge excitations in electron systems.
In addition, multipole transitions at high-momentum-transfer region open a new possibility to determine orbital-selective electronic excitations in multi-orbital itinerant 3$d$ electron systems.
As a theoretical example, we chose the antiferromagnetic state of iron arsenides and demonstrate that the orbital-selective excitations are detectable by choosing appropriate momentum transfer in NIXS.
We propose that both NIXS and resonant inelastic x-ray scattering are complementary to each other for fully understanding the nature of orbital excitations in multi-orbital itinerant electron systems.
\end{abstract}
\maketitle


\section{Introduction}

When the incident photon energy of inelastic x-ray scattering is far from the absorption edge, the scattering occurs without resonance enhancement.
This is called nonresonant x-ray scattering (NIXS).
NIXS has been used to measure both phonon and electronic excitations.
Among electronic excitations, charge excitations are capable by NIXS, since photons are predominantly scattered by charge.
Therefore, the dynamical density structure factor is detectable with NIXS,~\cite{VanHove1954,Sinha2001} and thus NIXS has some complementary aspects with inelastic neutron scattering.~\cite{Ishii2013}

The plane wave involved in the matrix element of NIXS can be expanded with respect to momentum transfer.
In the high momentum-transfer region, multipole transitions become relevant, and thus NIXS using high-energy x-ray has a unique capability of measuring higher multipole transitions.
By using this characteristic of NIXS, a possible observation of orbiton excitations in manganites has been proposed.~\cite{Ishihara2005}
Dipole-forbidden local $d$-$d$ excitations in Mott insulators, NiO and CoO, have successfully been observed via high momentum-transfer NIXS measurements.~\cite{Larson2007,Hiraoka2009}
The experimentally observed data have been analyzed theoretically on the basis of the first-principles calculations and cluster approaches.~\cite{Larson2007,Haverkort2007,Hiraoka2011}
 Furthermore, effective operators for the $d$-$d$ excitations have also established and their application to manganites has been proposed.~\cite{Veenendaal2008}

In itinerant 3$d$ electron systems, there is no clear localized $d$-$d$ excitation but corresponding excitations related to orbital degrees of freedom exist.
Therefore, it would be challenging to detect such orbital excitations by NIXS in high momentum-transfer region.
A good example of multi-orbital systems where the orbital degree of freedom is believed to be crucial for their physical properties would be iron-based superconductors.~\cite{Hosono2012}

In this paper, we demonstrate a possible realization of orbital-selective electronic excitations in the antiferromagnetic phase of iron arsenides by NIXS.
The antiferromagnetic phase is obtained by using the mean-field approximation for a five-orbital Hubbard model and dynamical susceptibilities are calculated within the random phase approximation (RPA).
We find that specific orbital-selective excitations are detectable by choosing appropriate momentum transfers.
Based on this finding, we propose that NIXS is a complementary experiment to resonant inelastic x-ray scattering (RIXS) tuned for Fe $L$ absorption edge, from which the orbital characters of excitations can be extracted.~\cite{Kaneshita2011}

The rest of the paper is organized as follows.
In Sec.~\ref{Sec2}, the model on the antiferromagnetic phase of iron-arsenide superconductors is introduced.
In Sec.~\ref{Sec3} the theory of NIXS is briefly introduced.
The matrix element of the plane wave with respect to Fe$^{2+}$ iron is shown in Sec.~\ref{Sec4}, together with the calculated NIXS spectra for BaFe$_2$As$_2$.
Finally, a summary is given in Sec.~\ref{Sec5}.

\section{Model Hamiltonian}
\label{Sec2}

We start with a multi-band Hubbard Hamiltonian for irons in iron-based superconductors given by $H_d = H_0 + H_I$.
The noninteracting Hamiltonian $H_0$ is given by
\begin{equation*}
H_0 = \sum_{i,j} \sum_{\sigma, \mu, \nu}  \left\{ t(\bm{\Delta}_{i, j}; \mu, \nu) + \varepsilon_\mu \delta_{\mu,\nu} \right\} c_{i,\mu,\sigma}^\dagger c_{j,\nu,\sigma},
\label{H0}
\end{equation*}
where $c_{i,\mu,\sigma}^\dagger$ creates an electron at site $i$ with orbital $\mu$ and spin $\sigma$.
$\bm{\Delta}_{i,j} \equiv \bm{r}_i-\bm{r}_j$, where $\bm{r}_i$ is the position of site $i$.
$\varepsilon_\mu$ and $t(\bm{\Delta}_{i,j};\mu,\nu)$ are the on-site energy and hopping integral, respectively.
The on-site energies and the hopping integrals are taken from Kuroki {\it et al.}~\cite{Kuroki2008}
The interaction Hamiltonian $H_I$ is expressed as~\cite{Oles1983}
\begin{eqnarray*}
H_I &=& U\sum_{i,\mu} c^\dagger_{i,\mu,\uparrow} c_{i,\mu,\uparrow} c^\dagger_{i,\mu,\downarrow} c_{i,\mu,\downarrow} \nonumber \\
 &+& (U-2J) \sum_{i,\mu \neq \nu} c^\dagger_{i,\mu,\uparrow} c_{i,\mu,\uparrow} c^\dagger_{i,\nu,\downarrow} c_{i,\nu,\downarrow} \nonumber \\
 &+& \frac{U-3J}{2} \sum_{i,\mu \neq \nu,\sigma} c^\dagger_{i,\mu,\sigma} c_{i,\mu,\sigma} c^\dagger_{i,\nu,\sigma} c_{i,\nu,\sigma} \nonumber \\
 &-& J \sum_{i,\mu \neq \nu} \left( c^\dagger_{i,\mu,\uparrow} c_{i,\mu,\downarrow} c^\dagger_{i,\nu,\downarrow} c_{i,\nu,\uparrow} - c^\dagger_{i,\mu,\uparrow} c_{i,\nu,\uparrow} c^\dagger_{i,\mu,\downarrow} c_{i,\nu,\downarrow} \right),
\label{HI}
\end{eqnarray*}
where $U$ is the intra-orbital Coulomb interaction, and $J$ is the Hund's coupling.
Here, we assume that the pair hopping is equal to $J$.

In order to describe an ordered phase, we construct the mean-field Hamiltonian $H_d^\mathrm{MF}$ from $H_d$ and self-consistently solve mean-field equations containing the order parameter defined by $\langle n_{\bm{Q},\mu,\nu,\sigma}\rangle = N^{-1}\sum_{\bm{k}} \langle c^\dagger_{\bm{k},\mu,\sigma} c_{\bm{k}+\bm{Q},\nu,\sigma} \rangle$ with the ordering vector $\bm{Q}$, where $N$ is the number of the lattice points, $c_{\bm{k},\mu,\sigma}^\dagger=\frac{1}{\sqrt{N}}\sum_i c_{i,\mu,\sigma}^\dagger e^{i\bm{k}\cdot\bm{r}_i}$, and the average $\langle\cdots\rangle$ is taken at zero temperature.
For an antiferromagnetically ordered state, we can rewrite the sum of the wave vectors as $\sum_{\bm{k}}\rightarrow\sum_{\bm{k}_0}\sum_{m=0,1}$ and $\bm{k}\rightarrow\bm{k}_0+m\bm{Q}$, where the sum of $\bm{k}_0$ is over the magnetically reduced Brillouin zone (BZ).

We consider a stripe-type antiferromagnetic order with $\bm{Q}=(\pi,0)$ observed in the parent compound of iron-arsenide superconductor, BaFe$_2$As$_2$.
We set $U=1.2$~eV and $J=0.22$~eV to yield a magnetic moment $m=0.8 \mu_B$ ($\mu_B$ is the Bohr magneton) at $n = 6.0$,~\cite{Kaneshita2011} where $n$ is the electron density and $m = \sum_\mu \langle n_{\bm{Q}, \mu, \mu, \uparrow} - n_{\bm{Q}, \mu, \mu, \downarrow} \rangle \mu_B$.
We note that $m$ is chosen to be close to the measured value in BaFe$_2$As$_2$.~\cite{Huang2008}

\section{Nonresonant Inelastic X-Ray Scattering}
\label{Sec3}

The spectral intensity of NIXS at zero temperature is given by
\begin{equation}
I(\bm{q},\omega)=-\frac{1}{\pi}\mathrm{Im} \langle\rho_{-\bm{q}}\frac{1}{\omega-H_d+E_0+i\delta}\rho_{\bm{q}}\rangle,
\label{I}
\end{equation}
where $E_0$ is the ground-state energy, $\delta$ is a small positive value.
By taking $\bm{q}=\bm{q}_\mathrm{r}+\bm{G}$ with the reciprocal vector, $\bm{G}$, and the reduced momentum in the reduced BZ, $\bm{q}_\mathrm{r}$, the operator $\rho_{\bm{q}}$ is defined as
\begin{equation}
\rho_{\bm{q}}=\sum_{\mu,\nu} \alpha_{\mu\nu}(\bm{q}_\mathrm{r}+\bm{G})\sum_{\bm{k},\sigma} c^\dagger_{\bm{k},\nu,\sigma} c_{\bm{k}+\bm{q}_\mathrm{r},\mu,\sigma}
\label{rho},
\end{equation}
with the coefficient $\alpha_{\mu\nu}(\bm{q})$ defined as
\begin{equation}
\alpha_{\mu\nu}(\bm{q}) \equiv \int \phi^*_\nu (\bm{r}) e^{i\bm{q}\cdot\bm{r}} \phi_\mu (\bm{r}) d\bm{r},
\label{MatElem}
\end{equation}
where $\phi_\mu (\bm{r})$ is the Wannier wave function of orbital $\mu$.
Using $\alpha_{\mu\nu}(\bm{q})$ and dynamical susceptibility $\chi^{\sigma\sigma'}_{\mu\nu\lambda\tau}(\bm{q}_\mathrm{r},\omega)$, the NIXS intensity reads
\begin{eqnarray}
I(\bm{q},\omega)&=&-\frac{1}{\pi} \mathrm{Im} \sum_{\mu,\nu}\sum_{\lambda,\tau} \alpha_{\mu\nu}(\bm{q}_\mathrm{r}+\bm{G}) \alpha_{\lambda\tau}(-\bm{q}_\mathrm{r}-\bm{G}) \nonumber \\
&\times& \sum_{\sigma,\sigma'} \chi^{\sigma\sigma'}_{\mu\nu\lambda\tau}(\bm{q}_\mathrm{r},\omega)
\label{Ic}
\end{eqnarray}
with
\begin{eqnarray}
\chi^{\sigma\sigma'}_{\mu\nu\lambda\tau}(\bm{q}_\mathrm{r},\omega)
&=& \sum_{\bm{k},\bm{k}'} \langle c^\dagger_{\bm{k}',\tau,\sigma'} c_{\bm{k}'-\bm{q}_\mathrm{r},\lambda,\sigma'} \nonumber \\
&\times& \frac{1}{\omega-H_d+E_0+i\delta}
 c^\dagger_{\bm{k},\nu,\sigma} c_{\bm{k}+\bm{q}_\mathrm{r},\mu,\sigma} \rangle.
\label{chi}
\end{eqnarray}
In this paper, the dynamical susceptibilities are calculated within RPA for the antiferromagnetically ordered phase given by $H_d^\mathrm{MF}$.~\cite{Kaneshita2011}

The plan wave $e^{i\bm{q}\cdot\bm{r}}$ in (\ref{MatElem}) can be expanded with respect to the partial waves,
\begin{equation}
e^{i\bm{q}\cdot\bm{r}}=\sum_{l=0}^\infty\sum_{m=-l}^{l} 4\pi i^l j_l(qr) Y_{lm}(\theta_q, \phi_q) Y_{lm}^*(\theta_r, \phi_r),
\label{Expansion}
\end{equation}
where $j_l(x)$ is the spherical Bessel function, $Y_{lm}(\theta,\phi)$ is the spherical harmonics, and $\theta_{q(r)}$ and $\phi_{q(r)}$ in $Y_{lm}$ represent polar angles of the $\bm{q}(\bm{r})$ vectors.
When $\bm{q}$ is located within the reduced BZ, i.e., $\bm{G}=0$, the $l=0$ component dominates the expansion (\ref{Expansion}).~\cite{Haverkort2007}
In this case, $\rho_{\bm{q}}$ (\ref{rho}) can be approximated by the charge operator given by
\begin{equation}
\rho_{\bm{q}}=N_{\bm{q}_\mathrm{r}}=\sum_{\bm{k},\mu,\sigma} c^\dagger_{\bm{k},\mu,\sigma} c_{\bm{k}+\bm{q}_\mathrm{r},\mu,\sigma}. 
\label{Nq}
\end{equation}
When $\bm{q}$ is larger than the reduced BZ, that is, $\bm{G}\neq 0$, non-zero $l$ contributes to the expansion (\ref{Expansion}).
This means that NIXS has a unique capability of measuring higher multipole transitions.

\begin{figure}
\includegraphics[width=0.3\textwidth]{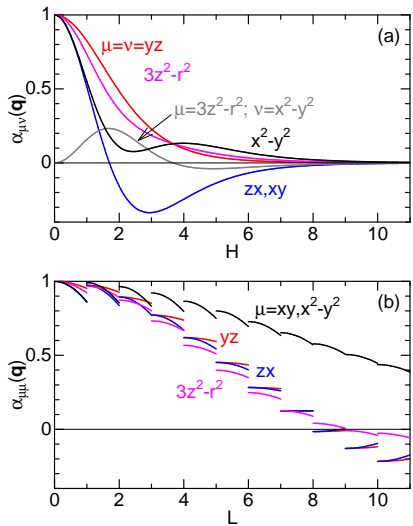}
\caption{(Color online)
The coefficient $\alpha_{\mu\nu}(\bm{q})$ defined in (\ref{MatElem}) for Fe$^{2+}$ ion. (a) $\bm{q}/(2\pi)=[H,0,0]$. (b) The orbital diagonal part of $\alpha_{\mu\mu}(\bm{q})$ along the $[0,0,L]$-$[1/2,0,L]$ direction for each value of $L$.}
\label{fig1}
\end{figure}

\section{Calculated Results and Discussions}
\label{Sec4}

In order to examine the $\bm{q}$ dependence of the coefficient (\ref{MatElem}) for iron arsenides, we simply use hydrogen-like atomic 3$d$ wave functions with the effective core charge $Z_\mathrm{eff}=8$ corresponding to Fe$^{2+}$ ion, instead of the Wannier wave functions.
We plot $\alpha_{\mu\nu}(\bm{q})$ along $\bm{q}/(2\pi)=[H,0,0]$ in Fig.~\ref{fig1}(a).
In a parent compound of iron-arsenide superconductor BaFe$_2$As$_2$ (Ba122), $[1,0,0]$ corresponds to $2\pi/a=2\pi/2.80=2.24$\AA$^{-1}$, where $a$ is the Fe-Fe bond length which is the half of the low-temperature orthorhombic lattice constant.
When $H$ is less than one, i.e., within the first BZ ($\bm{G}=0$), the diagonal components of $\alpha_{\mu\nu}(\bm{q})$ give almost the same value and there are very small off-diagonal components, leading to the charge operator (\ref{Nq}).
With increasing $H$, the diagonal components become unequal and off-diagonal one for $\mu=3z^2-r^2$ and $\nu=x^2-y^2$ emerges as shown in Fig.~\ref{fig1}(a).

\begin{figure}
\includegraphics[width=0.3\textwidth]{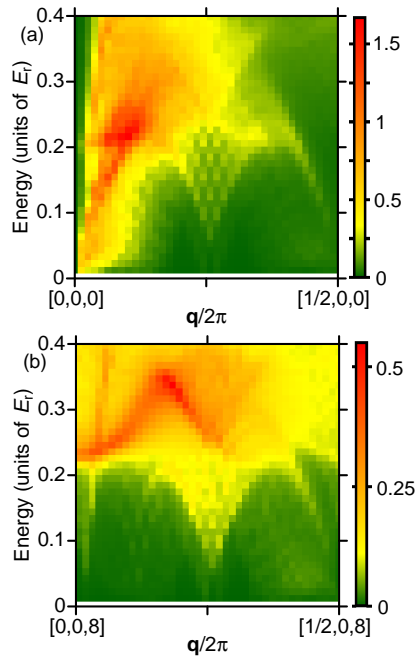}
\caption{(Color online)
Calculated NIXS spectra for the antiferromagnetic phase of Ba122. (a) Along the $[0,0,0]$-$[1/2,0,0]$ direction, and (b) along the $[0,0,8]$-$[1/2,0,8]$ direction. $E_\mathrm{r}$ is the energy unit introduced to take into account the renormalization of the band.
}
\label{fig2}
\end{figure}

Figure~\ref{fig2}(a) shows the calculated NIXS in the antiferromagnetic phase of Ba122 along $\bm{q}=\bm{q}_\mathrm{r}=2\pi [0,0,0]$ to $2\pi [1/2,0,0]$.
We note that the calculation has been performed for a two-dimensional model given by $H_d$.
In the figure, we use $E_\mathrm{r}$ as the unit of energy to evaluate the excitation energy under the presence of renormalization effects.
If there are no renormalization effects, $E_\mathrm{r}=1$~eV, but we guess that the value of $E_\mathrm{r}$ is around 0.5~eV as discussed separately.~\cite{Kaneshita2011}
The spectra shown in Fig.~\ref{fig2}(a) are almost equivalent to the dynamical charge susceptibility, and thus it is difficult to select orbital-resolved excitations.

Since our purpose is to predict orbital-selective excitations by NIXS, it is crucial for identifying the region of $\bm{q}$ where only a specific orbital contribution remains in $\alpha_{\mu\nu}(\bm{q})$.
After examining $\alpha_{\mu\nu}(\bm{q})$ in detail, we find such a $\bm{q}$ region where only two orbital diagonal components are dominant near $\bm{G}=2\pi [0,0,8]$.
To see this we show $\alpha_{\mu\mu}(\bm{q})$ along the $[0,0,L]$-$[1/2,0,L]$ direction for a given integer $L$ in Fig.~\ref{fig1}(b).
We note that in the interval 1 to 2 in the horizontal axis, for example, the diagonal components along $\bm{q}_\mathrm{r}/(2\pi)=[0,0,0]$ to $[1/2,0,0]$ with $\bm{G}/(2\pi)=[0,0,1]$ are plotted.
Near $L=8$, we can find that $\mu=xy$ and $\mu=x^2-y^2$ dominate $\alpha_{\mu\mu}(\bm{q})$.
This opens a possibility of observing orbital-selective excitations if one tunes $\bm{G}$ around this region.
We note that along this momentum-transfer direction the orbital off-diagonal components are very small as compared with diagonal ones.

\begin{figure}
\includegraphics[width=0.4\textwidth]{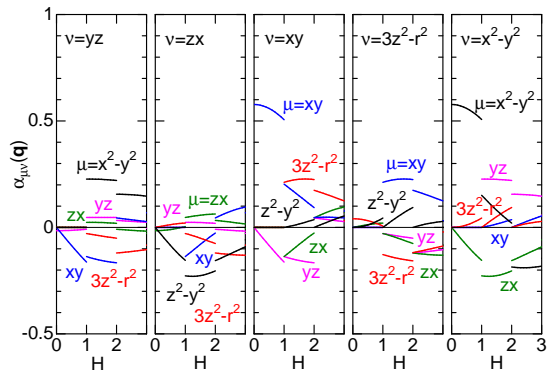}
\caption{(Color online)
The coefficient $\alpha_{\mu\nu}(\bm{q})$ defined in (\ref{MatElem}) for Fe$^{2+}$ ion. $\bm{q}/2\pi$ is along $[H,H,8]$-$[H+1/2,H,8]$. Each panel corresponds to each orbital $\nu$. 
}
\label{fig3}
\end{figure}

As a demonstration, we plot in Fig.~\ref{fig2}(b) calculated NIXS for Ba122 along $\bm{q}_\mathrm{r}/(2\pi)=[0,0,0]$ to $[1/2,0,0]$ starting from $\bm{G}/(2\pi)=[0,0,8]$.
We note again that the calculation has been done in the two-dimensional model and the $z$ component of $\bm{G}$ is taken into account through the matrix element $\alpha_{\mu\nu}(\bm{q})$.
There appears a momentum-dependent strong intensity around $\omega=0.3E_\mathrm{r}$, which is different from Fig.~\ref{fig1}(a) where all of orbitals can contribute to the intensity.
The intensity comes from either an $xy$ to $xy$ excitation or an $x^2-y^2$ to $x^2-y^2$ excitation.
Analyzing the wave functions in the ground state and corresponding excited states, we find that the strong intensity comes from the $xy$-$xy$ excitation.

In multi-orbital systems such as Ba122, there are characteristic orbital off-diagonal excitations, some of which have been predicted by the previous study by the same authors in connection with RIXS.~\cite{Kaneshita2011}
In order to detect such orbital off-diagonal excitations, we again need to identify $\bm{G}$ where the off-diagonal components are enhanced in $\alpha_{\mu\nu}(\bm{q})$.
We find that such a situation occurs along the $[H,H,8]$ direction.
Figure~\ref{fig3} exhibits $\alpha_{\mu\nu}(\bm{q})$ along $[H,H,8]$-$[H+1/2,H,8]$ for a given $H$.
When $\nu=3z^2-r^2$ and $H=1$, the magnitude of $\left| \alpha_{xy, 3z^2-r^2}(\bm{q}) \right|$ is as large as the diagonal contribution $\left| \alpha_{3z^2-r^2, 3z^2-r^2}(\bm{q}) \right|$.
Similarly, when $\nu=x^2-y^2$, the off-diagonal components $\alpha_{yz, x^2-y^2}(\bm{q})$ and $\alpha_{zx, x^2-y^2}(\bm{q})$ become large when $H=1$.
This implies possibility for the observation of orbital off-diagonal excitations in Ba122.
\begin{figure}
\includegraphics[width=0.3\textwidth]{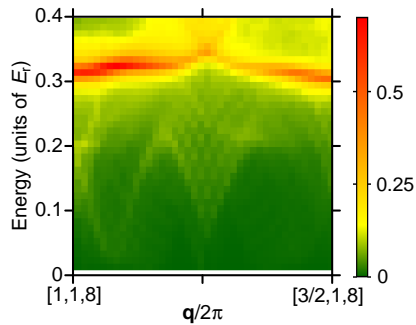}
\caption{(Color online)
Calculated NIXS spectra for antiferromagnetic phase of Ba122 along the $[1,1,8]$-$[3/2,1,8]$ direction. $E_\mathrm{r}$ is the energy unit introduced to take into account the renormalization of the band.
}
\label{fig4}
\end{figure}

In Fig.~\ref{fig4}, the NIXS intensity from  $[1,1,8]$ to $[3/2,1,8]$ ($H=1$ in Fig.~\ref{fig3}) is shown.
Strong intensity at $\omega=0.3E_\mathrm{r}$ is clearly shown, whose origin is attributed to excitations from $3z^2-r^2$ to $xy$ by the simulated RIXS.~\cite{Kaneshita2011}
Figure~\ref{fig4} is, thus, a prediction to NIXS experiment in the antiferromagnetic phase of Ba122 for detecting an orbital off-diagonal excitation from $3z^2-r^2$ to $xy$.
We note that the strong enhancement is a consequence of two contributions: one is the sharp occupied and empty peaks with dominantly $3z^2-r^2$ and $xy$ components, respectively, in the density of states~\cite{Kaneshita2010} and the other is the enhanced $\left| \alpha_{xy, 3z^2-r^2}(\bm{q}) \right|$ as shown in Fig.~\ref{fig3}.

\begin{figure}
\includegraphics[width=0.3\textwidth]{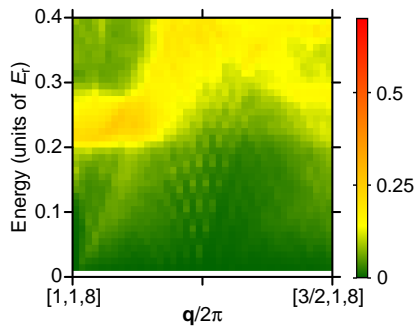}
\caption{(Color online)
The same as Fig.~\ref{fig4} but for the paramagnetic phase of Ba122.
}
\label{fig5}
\end{figure}

\begin{figure}
\includegraphics[width=0.3\textwidth]{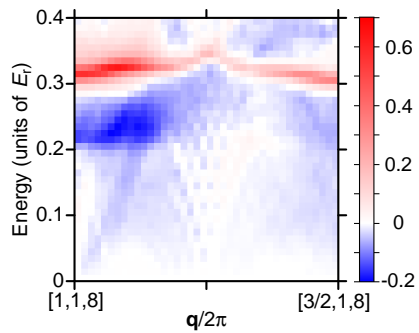}
\caption{(Color online)
The difference intensity of NIXS between the antiferromagnetic phase (Fig.~\ref{fig4}) and the paramagnetic phase (Fig.~\ref{fig5}).
}
\label{fig6}
\end{figure}

It is interesting to compare the NIXS intensity at the same $\bm{q}$ with that in the paramagnetic phase of Ba122.
In the paramagnetic phase, the calculated NIXS intensity is relatively broad as shown in Fig.~\ref{fig5} in contrast with the antiferromagnetic case in Fig.~\ref{fig4}.
The broadness predominantly comes from less structured density of states in the paramagnetic state.
As a result, the difference intensity shown in Fig.~\ref{fig6} enhances positive intensity around $\omega=0.3E_\mathrm{r}$, but it is negative below $\omega=0.3E_\mathrm{r}$.
We thus propose to measure the spectral change of NIXS across the N\'eel temperature in Ba122.
Although the scattering intensity calculated in the present study may be smaller than the phonon scattering by three or more orders of magnitude, as is the case of the $d$-$d$ excitations in NiO,~\cite{Baron2015} we expect that the spectral features proposed would be observed in the near future like the $d$-$d$ excitations in NiO.

\section{Summary}
\label{Sec5}
We have demonstrated theoretically orbital-selective electronic excitations in the antiferromagnetic phase of iron arsenides by employing NIXS.
The antiferromagnetic phase for Ba122 has been obtained by using the mean-field approximation for a five-orbital Hubbard model and dynamical susceptibilities have been calculated within RPA.
Using the formalism for NIXS, we have calculated its intensity for specific momentum-transfer regions where selected orbital contributions to NIXS are maximized.
We have clearly shown that the specific orbital-selective excitations are detectable by choosing appropriate momentum transfers.
For Ba122, we have predicted that the difference intensity along the specific momentum transfer picks up a specific orbital-selective excitation.
Based on this finding, we propose that NIXS and RIXS are complementary to each other for identifying orbital excitations in itinerant 3$d$ electron systems.

\begin{acknowledgments}
We thank A. Q. R. Baron for fruitful discussions.
This work was supported by a Grant-in-Aid for Scientific Research from the Japan Society for the Promotion of Science (Grants No. 24360036, No. 24540335, No. 26400381, No. 26287079, and No. 15H03553) and by MEXT HPCI Strategic Programs for Innovative Research (SPIRE) (hp150211) and Computational Materials Science Initiative (CMSI).
\end{acknowledgments}
\nocite{*}


\begin{thebibliography}{99}
\bibitem{VanHove1954}L. Van Hove, Phys. Rev. {\bf 95}, 249 (1954).
\bibitem{Sinha2001}S. K. Sinha, J. Phys.: Condens. Matter {\bf 13}, 7511 (2001).
\bibitem{Ishii2013}K. Ishii, T. Tohyama, and J. Mizuki, J. Phys. Soc. Jpn. {\bf 82}, 021015 (2013).
\bibitem{Ishihara2005}S. Ishihara, Y. Murakami, T. Inami, K. Ishii, J. Mizuki, K. Hirota, S. Maekawa, and Y. Endoh, New J. Phys. {\bf 7}, 119 (2005).
\bibitem{Larson2007}B. C. Larson, W. Ku, J. Z. Tischler, C.-C. Lee, O. D. Restrepo, A. G. Eguiluz, P. Zschack, and K. D. Finkelstein, Phys. Rev. Lett. {\bf 99}, 026401 (2007).
\bibitem{Hiraoka2009}N. Hiraoka, H. Okamura, H. Ishii, I. Jarrige, K. D. Tsuei, and Y. Q. Cai, Eur. Phys. J. B {\bf 70}, 157 (2009).
\bibitem{Haverkort2007}M. W. Haverkort, A. Tanaka, L. H. Tjeng, and G. A. Sawatzky, Phys. Rev. Lett. {\bf 99}, 257401 (2007).
\bibitem{Hiraoka2011}N. Hiraoka, M. Suzuki, K. D. Tsuei, H. Ishii, Y. Q. Cai, M. W. Haverkort, C. C. Lee, and W. Ku, Europhys. Lett. {\bf 96}, 37007 (2011).
\bibitem{Veenendaal2008}M. van Veenendaal and M. W. Haverkort, Phys. Rev. B {\bf 77}, 224107 (2008).
\bibitem{Hosono2012}"Special Issue on Iron-based Superconductors", edited by H. Hosono, H. Fukuyama, and H. Akai, Solid State Communi., Vol. 152, Issue 8, (2012), and references therein.
\bibitem{Kaneshita2011}E. Kaneshita, K. Tsutsui, and T. Tohyama, Phys. Rev. B {\bf 84}, 020511 (2011).
\bibitem{Kuroki2008}K. Kuroki, S. Onari, R. Arita, H. Usui, Y. Tanaka, H. Kontani, and H. Aoki, Phys. Rev. Lett. {\bf 101}, 087004 (2008).
\bibitem{Oles1983}A. M. Ole\'{s}, Phys. Rev. B {\bf 28}, 327 (1983).
\bibitem{Huang2008}Q. Huang, Y. Qiu, W. Bao, M. A. Green, J. W. Lynn, Y. C. Gasparovic, T. Wu, G. Wu, and X. H. Chen, Phys. Rev. Lett. {\bf 101}, 257003 (2008).
\bibitem{Kaneshita2010}E. Kaneshita and T. Tohyama, Phys. Rev. B {\bf 82}, 094441 (2010).
\bibitem{Baron2015}A. Q. R. Baron, arXiv:1504.01098 [cond-mat.mtrl-sci].

\end{thebibliography}

\end{document}